\documentclass[letterpaper, 10 pt, conference]{ieeeconf}
\IEEEoverridecommandlockouts
\usepackage{amsthm} 

\usepackage{longtable}
\usepackage{subfigure}
\usepackage{amsmath}
\allowdisplaybreaks
\usepackage{color}
\usepackage{footnote}
\usepackage{algorithm}
\usepackage{algpseudocode} 
\usepackage{algpseudocode}
\usepackage{algorithmicx}
\usepackage{multirow} 
\usepackage{booktabs}
\usepackage{graphicx}
\usepackage{amssymb}
\usepackage{amsbsy}
\usepackage{array}
\usepackage{longtable}
\usepackage{epstopdf}
\usepackage{pbox}
\usepackage{breqn}
\usepackage{mathrsfs}
\usepackage{multicol}
\usepackage{supertabular}
\usepackage{enumerate}
\usepackage{url}
\usepackage[justification=centering]{caption}

\newtheorem{thm}{Theorem}

\newtheorem{prp}{Proposition}

\newtheorem{lmm}{Lemma}

\begin{document}
\title{\LARGE \bf On Joint Convergence of Traffic State and Weight Vector in Learning-Based Dynamic Routing with Value Function Approximation}
\author{Yidan Wu, Jianan Zhang and Li Jin
\thanks{This work was in part supported by NSFC Project 62103260, SJTU UM Joint Institute, J. Wu \& J. Sun Foundation, and US NSF CMMI-1949710.}
\thanks{
Y. Wu and L. Jin are with the UM Joint Institute, Shanghai Jiao Tong  University, China. J. Zhang is with the School of Electronics, Peking University, China. (Emails: wyd510@sjtu.edu.cn, li.jin@sjtu.edu.cn, zhangjianan@pku.edu.cn)
}
}
\maketitle

\begin{abstract}
Learning-based approaches are increasingly popular for traffic control problems. However, these approaches are applied typically as black boxes with limited theoretical guarantees and interpretability. In this paper, we consider the theory of dynamic routing over parallel servers, a representative traffic control task, using semi-gradient on-policy control algorithm, a representative reinforcement learning method. We consider a linear value function approximation on an infinite state space; a Lyapunov function is also derived from the approximator. In particular, the structure of the  approximator naturally makes possible idling policies, which is an interesting and useful advantage over existing dynamic routing schemes. We show that the convergence of the approximation weights is coupled with the convergence of the traffic state. We show that if the system is stabilizable, then (i) the weight vector converges to a bounded region, and (ii) the traffic state is bounded in the mean. We also empirically show that the proposed algorithm is computationally efficient with an insignificant optimality gap.
\end{abstract}


\section{Introduction}
\label{sec:intro}

Dynamic routing is a classical control problem in transportation, manufacturing, and networking. This problem was conventionally challenging, because analytical characterization of the steady-state distributions of the traffic state and thus of the long-time performance metrics (e.g., queuing delay) are very difficult \cite{dai2022queueing,xie2022stabilizing}. Recently, there is a rapidly growing interest in applying reinforcement learning (RL) methods to dynamic routing and network control problems in general. RL methods are attractive because of their  computational efficiency  and adaptivity to unknown/non-stationary environments \cite{sutton2018reinforcement}. However, there is still a non-trivial gap between the demand for theoretical guarantees on key performance metrics and the black-box nature of RL methods. In particular, most existing theoretical results on RL are developed in the context of finite Markov decision processes (MDPs), while dynamic routing may be considered in infinite state spaces, especially for stability and throughput analysis.

\begin{figure}[htbp]
        \centerline{\includegraphics[width=0.8\linewidth]{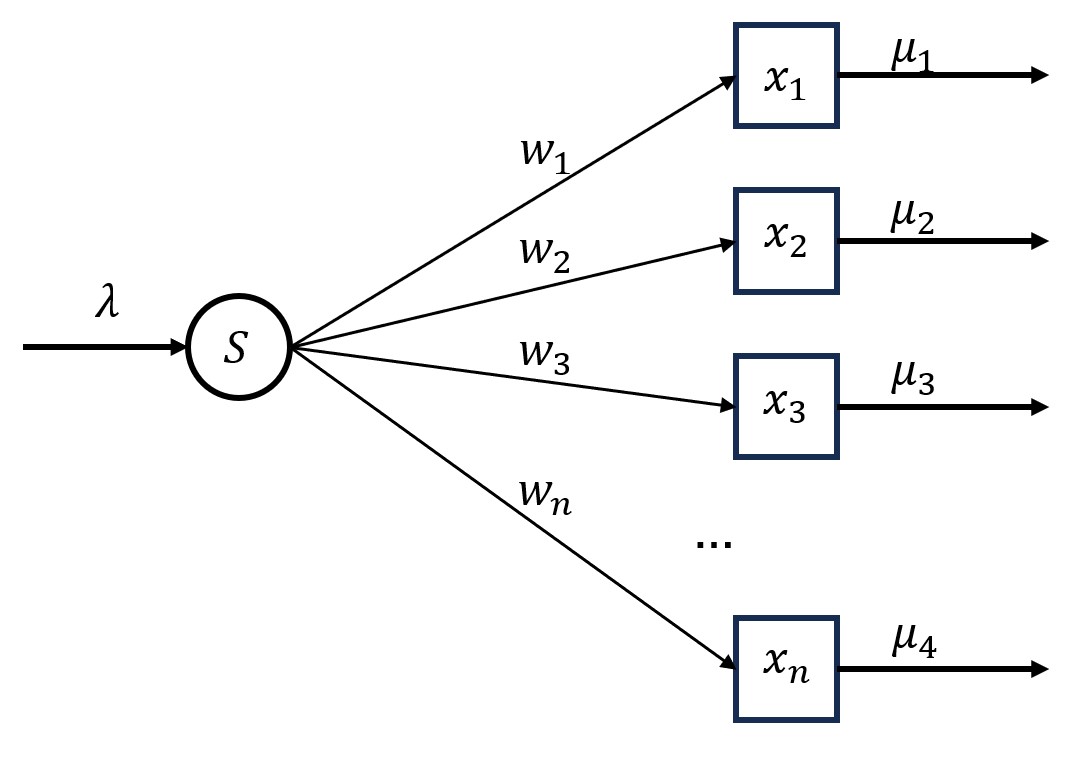}}
    \caption{A parallel service system.}
    \label{fig:parallelQN}
\end{figure}

In this paper, we make an effort to respond to the above challenge by studying the behavior of a parallel service system (Fig.~\ref{fig:parallelQN}) controlled by a class of semi-gradient SARSA (SGS) algorithms with linear function approximation; these methods are attractive because of (i) adaptivity to unknown model parameters and (ii) potential to obtain near-optimal policies.
Importantly, we jointly consider the convergence of the algorithm training process and of the traffic state process. The specific research questions are:
\begin{enumerate}
    \item How is the convergence of the weight vector coupled with the convergence of the traffic state?
    \item Under what conditions does the proposed SGS algorithm ensure the joint convergence of the weights and the state?
\end{enumerate}

The above questions involve two bodies of literature, viz. dynamic routing and reinforcement learning. 
Classical dynamic routing schemes rely on Lyapunov methods to study traffic stability and provide a solid foundation for our work \cite{kumar1995stability,dai1995stability}.
However, these methods are less powerful to search for routing policies that optimizing average system time. In particular, existing results usually assume non-idling policies, which may be quite restrictive.
Recently, RL is used for finding optimal routing policies and provides important tools and insights for practice \cite{bradtke1994reinforcement}.
In particular, Liu et al. proposed a mixed scheme that uses learning over a bounded set of states and uses a known stabilizing policy for the other states \cite{liu2022rl};
Xu et al. proposed a deep RL-based framework for traffic engineering problems in communication networks \cite{xu2018experience};
Lin et al. proposed an adaptive routing algorithm utilizing RL in a hierarchical network \cite{lin2016qos}.
However, existing theory on RL mostly, to the best of our knowledge, considers MDPs with finite or bounded state spaces \cite{gordon2000reinforcement,zhang2023convergence,de2000existence}; the theory on infinite/unbounded state spaces is limited to very special problems (e.g., linear-quadratic regulation \cite{lewis2013reinforcement}). 
Hence, existing learning-based routing algorithms either rely on empirical evidence for convergence or build on finite MDP theories. Thus, there is a lack of solid theory on the convergence of value function approximation over unbounded state spaces; such a theory is essential for developing interpretable and reliable routing schemes.

In response to the above research gaps, we jointly consider the convergence of traffic state and of the weight vector in dynamic routing. 
The traffic state characterizes the queuing process in the parallel service system illustrated in Fig.~\ref{fig:parallelQN}.
The routing objective is to minimize the expected total system time, which includes waiting times and service times.
The weights parameterize the approximate action value function, and thus determine the routing policy.
In particular, the algorithm naturally makes possible idling policies, which is an advantage over existing methods.
The weights are updated by a semi-gradient on-policy algorithm. 

Our main result (Theorem~\ref{thm: coupled stability}) states that the proposed algorithm ensures joint convergence of the traffic state and the weight vector if and only if the system is stabilizable. Importantly, we study the coupling between the long-time behavior of the traffic state and that of the weight vector, which extends the existing theory on finite-state RL \cite{zhang2023convergence} to unbounded state spaces. The convergence of traffic state results from a Lyapunov function associated with the approximate value function which verifies the drift criterion \cite{meyn_tweedie_1993}. The convergence of weights results from stochastic approximation theory \cite{tsitsiklis1996analysis}.
We compare the proposed algorithm with a much more sophisticated neural network-based algorithm and show that our algorithm converges much faster than the benchmark with only an 8\% optimality gap.

In summary, the contributions of this paper are as follows.
\begin{enumerate}
    \item We propose a structurally intuitive and technically sound algorithm to learn near-optimal routing policies over parallel servers. 

    \item We study joint convergence of traffic state and weight vector under the proposed algorithm; this is theoretically interesting in itself.

    \item We show empirical evidence for the computational efficiency and near-optimality of the proposed algorithm.
\end{enumerate}

The rest of this paper is organized as follows. Section~\ref{sec_model} introduces the parallel service system model, the MDP formulation, and the SGS algorithm. Section~\ref{sec_stable} presents and develops the main result on joint convergence. Section~\ref{sec_exp} compares the SGS algorithm with two benchmarks. Section~\ref{sec_con} gives the concluding remarks.
\section{Modeling and Formulation}
\label{sec_model}

Consider the system of parallel servers with infinite buffer sizes in Fig.~\ref{fig:parallelQN}. In this section, we model the dynamics of the system, formulate the dynamic routing problem as a Markov decision process (MDP), and introduce our semi-gradient SARSA (SGS) algorithm.
\subsection{System modeling}\label{subsec_SPQN}
Let $\mathcal{N}=\{1,2,3,\ldots,N\}$ be the set of parallel servers. Each server $n$ has an exponentially distributed service rate $\mu_n$ and job number $x_n(t)$ at time $t\in\mathbb{R}_{\geq 0}$. The state of the system is $x=[x_1,x_2,\ldots,x_N]^T$, and the state space is $\mathbb{Z}^N_{\geq 0}$. Jobs arrive at origin $S$ according to a Poisson process of rate $\lambda> 0$. When a job arrives, it will go to one of the $N$ servers according to a routing policy 
$$
\pi: \mathcal{N}\times \mathbb{Z}^N_{\geq 0}\rightarrow [0,1].
$$
That is, $\pi(a|x)$ is the probability of routing the new job to server $a$ conditional on state $x$. This paper focuses on a particular class of routing policies which we call the $weighted~shortest~queue$ (WSQ) policy. WSQ is based on the approximation for the action value function $\hat Q: \mathcal{N}\times \mathbb{Z}^N_{\geq 0}\times \mathbb{R}^N_{> 0}\to \mathbb{R}_{\geq 0}$ defined as:
\begin{align}
\label{equa:def Q}
    \hat Q(x,a;w) := \sum_{n=1}^N w_n(x_n+\mathbb{I}_{\{n=a\}})^2,
\end{align}
where $w=[w_1,w_2,\ldots,w_N]^T \in\mathbb{R}^N_{> 0}$ is the weight vector. For technical convenience, we consider the softmax version of WSQ  
\begin{align}
\label{equa: policy}
    \pi_{w}(a|x)=\frac{\exp(-\hat Q(x,a;w)/\iota)}{\sum_{b=1}^{N}\exp(-\hat Q(x,b,w)/\iota)},
\end{align}
where $\iota\in(0,\infty)$ is the temperature of the softmax function. Note that $\pi_w(a|x)$ converges to a deterministic policy greedy w.r.t. $\hat Q$ as $\iota$ approaches 0 \cite{zhang2023convergence}.

We say that the traffic in the system is \textit{stable} if there exists a constant $M<\infty$ such that for any initial condition,
\begin{align}
    \label{equa: stable state}
    \lim_{t\rightarrow\infty}\frac{1}{t}\int_{s=0}^t\mathsf E[\|x(s)\|_1]ds<M.
\end{align}
We say that the system is \textit{stabilizable} if 
\begin{align}
    \label{equa: stabilizable}
    \lambda<\sum_{n=1}^N\mu_n.
\end{align}
Note that the above ensures the existence of at least a stabilizing Bernoulli routing policy. 

\subsection{MDP formulation}
Since routing actions are made only at transition \textit{epochs} \cite[p.72]{gallager2013stochastic}, the routing problem of the parallel queuing system can be formulated as a discrete-time (DT) MDP with countably infinite state space $\mathbb{Z}^N_{\geq 0}$ and finite action space $\mathcal{N}$. With a slight abuse of notation, we denote the state and action of the DT MDP as $x[k]\in\mathbb{Z}^N_{\geq 0}$ and $a[k]\in\mathcal{N}$, respectively. Specifically, $x[k]=x(t_k)$, where $t_k$ is the $k$-th transition epoch of the continuous-time process. As indicated in Section~\ref{subsec_SPQN}, the routing policy can be parameterized via weight vector $w\in\mathbb{R}^N_{>0}$. 

The transition probability $\mathsf p(x'|x,a)$ of the DT MDP can be derived from the system dynamics in a straightforward manner. Let $e_i\in\{0,1\}^N$ denote the unit vector such that $e_{i,i}=1$ and $e_{i,j}=0,j\not=i$. Then we have
\begin{align*}
    \mathsf p(x'|x,a)=\begin{cases}
        \frac{\lambda}{\lambda+\sum_{n=1}^N\mu_n\mathbb{I}_{\{x_n>0\}}} &x'\in\{x+e_a\}^N_{a=1},\\
        \frac{\mu_n\mathbb{I}_{\{x_n>0\}}}{\lambda+\sum_{n=1}^N\mu_n\mathbb{I}_{\{x_n>0\}}}&x'=x-e_n.
    \end{cases}
\end{align*}

The one-step random cost of the MDP is given by 
\begin{align*}
    c[k+1]=&\|x[k+1]\|_1(t_{k+1}-t_k),
\end{align*}
where $\|\cdot\|_1$ is the 1-norm for $\mathbb{R}^N$. Let $\bar c(x,a)=\mathsf E[c[k+1]|x[k]=x,a[k]=a]$ denote the expected value of cost.
The total discounted cost over infinite-horizon process is thus given by
\begin{align*}
    Q_\pi(x,a)=&\mathsf{E}_\pi\Big[\sum^\infty_{\ell=0}\gamma^{\ell}\|x[\ell+1]\|_1(t_{\ell+1}-t_{\ell})\Big|x,a\Big],
\end{align*}
where $\gamma\in(0,1)$ is the discount factor of infinite-horizon MDP. Let $Q^*(x,a)$ denote the solution of Bellman optimal equation, that
\begin{align*}
    Q^*(x,a)=\bar c(x,a)+\gamma\min_{a'}\sum_{x'}\mathsf p(x'|x,a)Q^*(x',a'),
\end{align*}
and let $\pi^*$ denote the greedy policy with respect to $Q^*$.

Closed-form solution to $Q_\pi$ is not easy. Hence, we use the $\hat Q$ function defined by \eqref{equa:def Q} as a proxy for $Q_\pi$. Motivated by \cite{zhang2023convergence,tsitsiklis1996analysis}, we consider the function approximator as
\begin{align}
    \label{equa:def V}
    &\hat Q(x,a;w)=\sum_{n=1}^N w_n\phi_n(x,a),\\
    &\phi_n(x,a)=(x_n+\mathbb{I}_{\{n=a\}})^2,\quad n=1,2,\ldots,N,\nonumber
\end{align}
where $\phi_n:\mathcal{N}\times\mathbb{Z}^N_{\geq 0}\to\mathbb{R}_{\geq0}$ and $\{\{\phi_n\}_{x,a}\}_{n=1}^N$ are linearly independent basis functions. Let $w^*$ denote the optimal solution to
\begin{align*}
    \min_w\sum^N_{\substack{x\in\mathbb{Z}^N_{\geq0},\\ a\leq N}}d^*(x)\pi^*(a|x)\Big(Q^*(x,a)-\hat Q(x,a;w)\Big)^2,
\end{align*}
where $d^*(x)$ is the invariant state distribution under policy $\pi^*$. We select quadratic basis functions because $Q_\pi$ is non-linear and the quadratic function is one of the simplest non-linear functions. Besides, the analysis is generalizable to polynomials with higher order. 

\subsection{Semi-gradient SARSA algorithm}
\label{subsec: sgs algorithm}
Inspired by \cite{zhang2023convergence}, let $w[k]$ denote the weight vector at the $k$-th transition epoch, which is updated by an SARSA(0) algorithm
\begin{align*}
    &w[k+1]=\Gamma\Big(w[k]+\alpha_k\Delta[k]\nabla_w\hat Q(x[k],a[k];w[k])\Big);
\end{align*}
in the above, $\Gamma:\mathbb{R}_{>0}^N\to\mathbb{R}_{>0}^N$ is a projection operator, $\alpha_k$ is the stochastic step size, $\Delta[k]$ is the temporal-difference (TD) error, and $\nabla_w\hat Q(x[k],a[k];w[k])$ is the gradient, which are specified as follows.

The projection $\Gamma(\cdot)$ is defined with a positive constant $C_\Gamma$:
\begin{align*}
\Gamma(w)=\begin{cases}
    w&\|w\|\leq C_\Gamma,\\
    C_\Gamma\frac{w}{\|w\|} &\|w\|> C_\Gamma,
\end{cases}
\end{align*}
where $\|\cdot\|$ is the standard $2$-norm. Besides, we use $\langle x,y \rangle := x^Ty$ denote the standard inner product in Euclidean spaces. 

The temporal difference (TD) error $\Delta[k]$ and the gradient $\nabla_w\hat Q(x[k],a[k];w[k])$ are as follows.
Let $\phi=[\phi_1,\phi_2,\ldots,\phi_N]^T$, then we can compactly write
\begin{align*}
    &\hat Q(x,a;w)=\phi^T(x,a)w,\\
    &\triangledown_w\hat Q(x,a;w)=\phi(x,a).
\end{align*}
Then, for any $k\in\mathbb{Z}_{\geq 0}$, the TD error and the gradient are collectively given by
\begin{align*}
    &\delta_{w[k]}(x[k],w[k])
    =\Delta[k]\nabla_w\hat Q(x[k],a[k];w[k])\\
    &=
    \Bigg(-\phi^T\Big(x[k],a[k]\Big)w[k]+c[k+1]\\
    &\qquad+\gamma\phi^T\Big(x[k+1],a[k+1]\Big)w[k]\Bigg)\phi\Big(x[k],a[k]\Big).
\end{align*}

The step sizes ${\{\alpha_k\}}$ are generated by the following mechanism. We define an auxiliary sequence $\{\tilde{\alpha}_{\tilde{k}};\tilde{k}\in\mathbb{Z}_{\geq 0}^N\}$ satisfying the standard step size condition \cite{gordon2000reinforcement}
\begin{align}
    \sum_{k=0}^\infty\tilde{\alpha}_{\tilde{k}}=\infty,
    \quad
    \sum_{k=0}^\infty\tilde{\alpha}^2_{\tilde{k}}<\infty.
    \label{eq_sumk=0}
\end{align}
The step sizes can not be too small to stop the iteration process, while also can not be too large to impede convergence. Let $B_\alpha$ denote a finite positive constant and define 
$$
\tilde{k}=\max\nolimits_{\substack{\delta_{w[k]}(x[k],w[k])\leq B_\alpha \\ k'<k}} k'.
$$
Then the step size sequence $\{\alpha_k\}$ can be constructed as
\begin{align*}
    \alpha_k=\begin{cases}
        \tilde{\alpha}_{\tilde{k}}&\delta_{w[k]}(x[k],w[k])\leq B_\alpha,\\
        0&o.w.,
    \end{cases}
\end{align*}
where $B_\alpha$ is a finite positive constant.   
That is, $\{\alpha_k\}$ consists of zeros and elements from the deterministic sequence $\{\tilde{\alpha}_{\tilde{k}}\}$ as demonstrated in Table~\ref{table:stepsize}. Thus the weight vector $w[k+1]$ is updated only when the constraint $B_\alpha$ is satisfied.
\begin{table}[htbp]
    \centering
    \caption{A sample of stochastic step sizes $\{\alpha_k\}$.}
    \label{table:stepsize}
    \begin{tabular}{cccc}
        \toprule  
        $k$&$\delta_{w[k]}(x[k],w[k])\leq B_\alpha?$&$\alpha_k=$\\
        \midrule  
        0&Yes&$\tilde{\alpha}_{0}$\\
        1&Yes&$\tilde{\alpha}_{1}$\\
        2&No&$0$\\
        3&Yes&$\tilde{\alpha}_{2}$\\
        $\vdots$&$\vdots$&$\vdots$\\
        \bottomrule  
    \end{tabular}
\end{table}

The update equation thus becomes
\begin{align}
    w[k+1]=\Gamma\Big(w[k]+\alpha_k\delta_{w[k]}(x[k],w[k])\Big).
    \label{equa: update w SARSA0}
\end{align}
It is known that SARSA chatters when combined with linear function approximation \cite{gordon2000reinforcement}. We say that Algorithm~\ref{algo: SGS-WSQP} is \textit{convergent to a bounded region} if there exists a positive finite constant $B$ such that
\begin{align}
\label{equa: algo convergent}
    \lim_{k\rightarrow\infty}\mathsf E[\|w[k]-w^*\|]\leq B,
\end{align}
for every initial traffic state $x[0]\in\mathbb{Z}_{\geq 0}^N$ and every initial weight $w[0]\in\mathbb{R}^N_{> 0}$.
\begin{algorithm}
\caption{(SGS) Computation of $\hat Q$ for $Q$}
\label{algo: SGS-WSQP}   
\begin{algorithmic}[1]
  \Require Initial weights $w[0],\|w[0]\|<C_\Gamma$, WSQ policy $\pi_{w[0]}$, step sizes sequence $\alpha_k$, $\gamma$
  \State Initialize weights $w[0]\leftarrow w[0]$
  \For{$k=0,1,\ldots$}
    \State Execute action $a[k]$
    \State Obtain new state $x[k+1]$ and immediate reward $c[k+1]$
    \State Select $a[k+1]$ according to policy $\pi_{w[k]}$
    \State Calculate $\delta_{w[k]}(x[k],w[k])$
    \State $w[k+1] \leftarrow  \Gamma(w[k]+\alpha_k\cdot\delta_{w[k]}(x[k],w[k])) $
  \EndFor
\end{algorithmic}
\end{algorithm}
\section{Joint convergence guarantee}
\label{sec_stable}
In this section, we develop the main result of this paper, which states that the proposed semi-gradient SARSA (SGS) algorithm ensures joint convergence of traffic state and weight vector if and only if the parallel service system is stabilizable.

\begin{thm}(Joint convergence)
\label{thm: coupled stability}
Consider a stabilizable parallel service system with arrival rate $\lambda>0$ and service rates $\mu_1,\mu_2,\ldots,\mu_N>0.$ Suppose that the step size condition \eqref{eq_sumk=0} holds. Then, the traffic state $x[k]$ converges in the sense of \eqref{equa: stable state} and the weight $w[k]$ converges in the sense of \eqref{equa: algo convergent}.
\end{thm}

The above result essentially states that the joint convergence of $w[k]$ and $x[k]$ relies on step size constraint \eqref{eq_sumk=0}. They are standard for reinforcement learning methods, which ensures that (i) sufficient updates will be made, and (ii) randomness will not perturb the weights at steady state \cite{tsitsiklis1996analysis}. 

The rest of this section is devoted to the proof of the above theorem. Section \ref{subsec: stability of SPQN} proves the stability of traffic state, section \ref{subsec: premises of SGS} presents unboundedness of $\sum_{k=0}^\infty \alpha_k$, and section \ref{subsec: convergence of SGS} proves the convergence of the approximation weights.

\subsection{Convergence of $x[k]$}
\label{subsec: stability of SPQN}
In this section we construct a Lyapunov function and argue the drift to show the convergence of traffic state $x[k]$, with policy $\pi$ and proper temperature parameter $\iota$. In particular, we considering the Lyapunov function $$\hat V_w(x)=\sum_{n=1}^Nw_nx_n^2$$ with the same weight vector as \eqref{equa:def V}. We show that there exist $\iota>0,~\epsilon_v > 0,~B_v < \infty$ such that for all $x \in \mathbb{Z}^N_{\geq 0}$ and for all $w\in\mathbb{R}^N_{> 0}$
\begin{align}
    \label{equa: drift of V}
    \mathcal{L}\hat V_w({x})=-\epsilon_v \sum_{n=1}^Nw_nx_n + B_v,
\end{align}
where $\mathcal L$ is the infinitesimal generator of the system under the (softmax) WSQ policy.

Let $l_n=\mathbb{I}_{\{x_n\geq 1\}}$, $m=\arg\min_{n\leq N}w_n(2x_n+1)$. We have 
\begin{align*}
    \mathcal{L} \hat V_w(x)&=\sum\nolimits_{n=1}^Nl_n\mu_nw_n(-2x_n+1)\\
    &+\sum\nolimits_{a=1}^N\pi_w(a|x)\lambda_{a}w_{a}(2x_{a}+1).
\end{align*}
Suppose that $\iota$ is sufficiently small, since $(l_n-1)x_n=0$, we have
\begin{align*}
    \mathcal{L} \hat V_w(x)\leq\sum_{n=1}^N\Big(\lambda\frac{\mu_n}{\sum_{k=1}^N \mu_k}-\mu_n\Big)w_n(2x_n+1)+B_v.
\end{align*}
Then \eqref{equa: drift of V} holds when \eqref{equa: stabilizable} holds. We can use Foster-Lyapunov criterion \cite[Theorem 4.3.]{meyn_tweedie_1993} conclude \eqref{equa: stable state}.
\subsection{Unboundedness of $\sum_{k=0}^\infty\alpha_k$}

\label{subsec: premises of SGS}
In this section, we show that $\sum_{k=1}^\infty \alpha_k=\infty$ a.s..
\begin{lmm}
    \label{lmm: A bounded}
    Under the constraints in Theorem~\ref{thm: coupled stability}, let $W_p(x)=\sum_{n=1}^Ne^{\nu w_n (2x_n+1)}/w_n,~\nu>0$, then there exists function $g_p\geq1$ and finite non-negative constant $B_w$ satisfying 
    \begin{align}
    \label{equa: W_p drift condition}
        \Delta W_p(x) &= \mathsf{E}[W_p(x[k+1])-W_p(x[k])|x[k]=x] \nonumber\\
        &\leq-g_p(x)+B_w.
    \end{align}
    Furthermore, we have
    \begin{align*}
        \lim_{K\rightarrow\infty}\frac{1}{K}\sum\nolimits_{k=0}^{K-1}\mathsf E[g_p(x)]\leq B_w.
    \end{align*}
\end{lmm}
\noindent\emph{Proof:}
    Considering similar definition of $m,l_n$ in section~\ref{subsec: stability of SPQN}. Similarly, under the (softmax) WSQ policy, we have
    \begin{align*}
        &\Delta W_{p}(x) = \sum_{a=1}^N\pi_w(a|x)\Big[\sum\limits_{n\not=a}^N\frac{l_n}{w_n}\cdot\Big(e^{\nu w_n(2x_n-2_mu_n+1)}\\
        &\quad-e^{\nu w_n(2x_n+1)}\Big)-\frac{1}{w_a}\Big((1-l_{a})\cdot e^{\nu w_a(2x_a+2\lambda+1)}\\
        &\quad+ l_{a}\cdot e^{\nu w_a(2x_a+2(\lambda-\mu_a)+1)}+e^{\nu w_a(2x_a+1)}\Big)\Big].
    \end{align*}
    Note that there is $l_n\cdot e^{\nu w_n2x_n} = (l_n-1) + e^{\nu w_n2x_n}$ and suppose that $\iota$ is sufficiently small, we have
    \begin{align}
    \label{equa: W_p equality}
        &\Delta W_{p}(x) = \sum_{n\not=m}^Ne^{\nu w_n(2x_n+1)}\cdot(e^{-2\nu w_n\mu_n}-1)/w_n\nonumber\\
        &+e^{\nu w_m(2x_m+1)}\cdot(e^{2\nu w_{m}(\lambda-\mu_{m})}-1)/w_m+B_{0},
    \end{align}
    where $B_{0}=\frac{1}{w_m}(e^{\nu w_m(2x_m+1)}-e^{\nu w_{m}(2\lambda_{m}-2\mu_{m}+1)})+\sum_{n\not=m}^Ne^{\nu w_n}(1-e^{-2\nu w_n\mu_n}))\frac{1}{w_n}$ is a finite positive constant. In the case of $\lambda\leq\mu_{m}$, the drift equation \eqref{equa: W_p equality} naturally satisfies \eqref{equa: W_p drift condition}. When $\lambda>\mu_m$, we have 
    \begin{align*}
    &\Delta W_p(x)\leq\sum_{n\not =m}^Ne^{\nu w_n(2x_n+1)}\cdot\Lambda(\nu),\\
    &\Lambda(\nu)=\frac{\mu_n(e^{2\nu w_m(\lambda-\mu_m)}-1)}{w_m\sum_{k\not= m}^N \mu_k}+\frac{(e^{-2\nu w_n\mu_n}-1)}{w_n}.
    \end{align*}
    Note that $\Lambda(0)=0,~\Lambda(\infty)\rightarrow{\infty}$. The derivate of $\Lambda(\nu)$ at $\nu=0$ is calculated as
    \begin{align*}
        &\frac{d\Lambda}{d\nu}\Bigr|_{\nu=0} =2\mu_n(\frac{\lambda-\mu_m}{\sum_{k\not =m}^N \mu_k}-1).
    \end{align*}
    Then the derivative of $\Lambda$ is negative when $\lambda<\sum_{n=1}^N\mu_n$, which implies that there exist $\nu_0>0$ as the second zero of $\Lambda(\nu)$ and $\Lambda(\nu)<0,~\nu\in(0,\nu_0)$. Now we can conclude that with a proper selection of $\nu$, \eqref{equa: W_p drift condition} is guaranteed. There exists a finite positive constant $B_p$ satisfies
    $$
    g_p(x)=B_p\sum\nolimits_{n=1}^Ne^{\nu w_n(2x_n+1)}/w_n+1.
    $$
    Following the proof in \cite{georgiadis2006resource}, summing the inequality over epochs $k\in\{0,\ldots,K-1\}$ yields a telescoping series on the left hand side of \eqref{equa: W_p drift condition}, result in
    \begin{align*}
        &\mathsf{E}[W_p(x[K])]-\mathsf{E}[W_p(x[0])]\\
        &\leq K(B_0+1)-\sum_{k=0}^{K-1} \mathsf E[B_p\sum\nolimits_{n=1}^Ne^{\nu w_n(2x_n+1)}/w_n+1].
     \end{align*}
    Since $\mathrm{E}[W_p(x[0])]\geq0$, we have
    \begin{align*}
        \lim_{K\rightarrow\infty}\frac{1}{K}\sum_{k=0}^{K-1} \mathsf E[B_p\sum\nolimits_{n=1}^Ne^{\nu w_n(2x_n+1)}/w_n+1]\leq B_w,
    \end{align*}
    where $B_w=B_0+1$. The above inequality implies the boundedness of $\mathsf E[\sum_{n=1}^Ne^{\nu w_n(2x_n+1)}]$, thus the higher order stability of system states \cite{meyn2008control}.
\qed
\begin{prp}
\label{prp:ergodic}
    $\forall w\in\mathbb{R}^N_{>0}$, the chain induced by $\pi_w$ is ergodic and positive Harris recurrent.
\end{prp}
\noindent\emph{Proof:}
    To argue for the irreducibility of the chain, note that the state $x=\mathbf{0}$ can be accessible from any initial condition with positive probability. According to \cite[Theorem 11.3.4]{meyn2012markov}, the proof of ergodic and positive Harris recurrent is straightforward with Lemma~\ref{lmm: A bounded}. 
\qed
\begin{prp}
\label{prp:alpha}
    With Proposition~\ref{prp:ergodic}, the step size sequence $\{a_k\}$ in SGS satisfies \eqref{eq_sumk=0}.
\end{prp}
\noindent\emph{Proof:}   
    Let $\tilde{\mathcal{X}}:=\{x:\|x\|\leq B_{\tilde{\mathcal{X}}},x\in\mathbb{Z}_{\geq 0}\}$, where $B_{\tilde{\mathcal{X}}}$ is a finite positive constant and there is $\tilde{\mathcal{X}}\in\mathbb{Z}^n_{\geq 0}$. Then the boundedness $\Delta[k]\nabla_w\hat Q(a[k],x[k],w[k])\leq B_\alpha$ can be satisfied by constraining $x[k]\in\tilde{\mathcal{X}}$ and $t_{k+1}-t_k\leq B_T$, where $B_T<\infty$ is a positive constant. That is, the weight vector $w[k+1]$ is updated only when the state and time interval (i.e., the immediate cost) are not too large.
    
    Let use $\tau_{\tilde{\mathcal{X}}}:=\sum_{k=1}^\infty\mathbb{I}_{x[k]\in\tilde{\mathcal{X}}}$ denote the occupation time of states $x\in\tilde{\mathcal{X}}$, since the chain is positive Harris recurrent, we have $P(\tau_{\tilde{\mathcal{X}}}=\infty)=1$. Since the arrival rate and service rates are well-defined, we have $P(t_{k+1}-t_k\leq B_T)\geq P_T>0$. Then we have
    \begin{align*}
    &\sum_{k=0}^\infty\alpha_k\geq P_T\sum_{x_k\in\tilde{\mathcal{X}}}^{\tau_{\tilde{\mathcal{X}}}}\tilde{\alpha}_{\tilde{k}}+0=\infty\quad \text{a.s.},\\
    &\sum_{k=0}^\infty\alpha_k^2\leq\sum_{k=0}^\infty\tilde{\alpha}^2_{\tilde{k}}<\infty\quad\text{a.s.}
    \end{align*}
    as desired.
\qed
\subsection{Convergence of w[k]}
\label{subsec: convergence of SGS}
In the following, we establish the convergence of $w[k]$ by showing (i) the convergence under fixed-policy evaluation and (ii) the difference among optimal weight vectors due to policy improvement is bounded.

\begin{prp}(Lipschitz continuity of $\pi_w(a|x)$) 
\label{prp:Lip pi} For our WSQ policy, there exists $L_\pi>0$ such that $\forall w,w',a,x$,
$$
\|\pi_w(a|x)-\pi_{w'}(a|x)\|\leq L_{\pi}\|w-w'\|.
$$
\end{prp}
\noindent\emph{Proof:}
    Note that the boundedness of derivative towards $w$ implies the Lipschitz continuity. With the well constructed sequence $\{\alpha_k\}$, we have 
    $$
    |\pi_w(a|x)-\pi_{w'}(a|x)|=0= L_{\pi}\|w-w'\|, \quad x\not\in\tilde{\mathcal{X}}.
    $$
    For $x\in\tilde{\mathcal{X}}$, according to \eqref{equa: policy}, we have $\pi_w(a|x)\in(0,1)$ and
    \begin{align*}
    \Big|\frac{d\pi_w(a|x)}{dw}\Big| \leq\Big|\frac{N}{\iota}\max_{a,b\leq N}2\cdot|x_a-x_b|\Big|,
    \end{align*} which is bounded.
\qed

With the above Proposition~\ref{prp:ergodic}-\ref{prp:Lip pi}, we can analysis the fixed policy performance and bound the difference among distinct policies. For a better elucidation, we use subscript $w^*[k]$ denote the invariant steady-state dynamic of the chain under fixed policy $\pi_{w[k]}$, and use subscript $k$ denote the real dynamic at the $k$-th transition. That is $d_{w^*[k]}(\cdot)$ denotes the invariant distribution of policy $\pi_{w[k]}$, and $d_k(\cdot)$ denotes the state distribution at the $k$-th transition. Suppose that under the fixed policy $\pi_{w[k]}$ and according to \cite{melo2008analysis}, we have
\begin{align*}
    \bar \delta_{w[k]}(x[k],w[k]) =A_{w^*[k]}w[k]+b_{w^*[k]},
\end{align*}
where $A_{w^*[k]}$ is defined as
\begin{align*}
    &A_{w^*[k]}\\
    &=\sum_{a'\leq N,\atop x'\in\mathbb{Z}^N_{\geq 0}}\Big(\sum_{a\leq N,\atop x\in\mathbb{Z}^N_{\geq 0}}d_{w^*[k]}(x)\pi_{w[k]}(a|x)\mathsf p_{w^*[k]}(x'|a,x)\\
    &\times\gamma\phi(x,a)-d_{w^*[k]}(x')\phi(x',a')\Big)\pi_{w[k]}(a'|x')\phi^T(x',a')\\
    &\leq(\gamma-1)\mathsf E_{w^*[k]}\Big[\phi(x',a')\phi^T(x',a')\Big],
\end{align*}
Which is negative definite since $\gamma\in(0,1)$. Analogous, we have $b_{w^*[k]}=\mathsf E_{w^*[k]}[\phi^T(x,a)c],b_k=\mathsf E_k[\phi^T(x,a)c],A_k=\mathsf E_k[\phi(x,a)(\gamma\phi^T(x',a')-\phi^T(x,a))]$ and $\mathsf E[\delta_{w[k]}(x,w)]=A_kw+b_k$.

According to \cite[Theorem 14.0.1]{meyn2012markov}, we have
\begin{align}
\label{equa:bound A}
    \sup_{f_p:|f_p|\leq g_p}&\sum_{t=0}^\infty\Big|\sum_{x'}\mathsf p_k^t(x'|x)f_p(x')-\sum_{x'}d_{w^*[k]}(x')f_p(x')\Big|\nonumber\\
    &<B_f(W_p(x)+1),\quad x[k]=x,x\in\tilde{\mathcal{X}},
\end{align}
where $B_f$ is a finite positive constant. $\mathsf p^t_k(x'|x)$ indicates the transition probability from state $x$ to $x'$ after $t$ steps under policy $\pi_{w[k]}$, and there is $\mathsf p(x'|x)=\sum_{a=1}^N\pi(a|x)\mathsf p(x'|a,x)$. According to \cite{tsitsiklis1996analysis}, with \eqref{equa:bound A} holds, the iterative algorithm \eqref{equa: update w SARSA0} has a unique solution $w^*$ satisfies that $\bar\delta(x,w^*)=0$ under fixed policy. With a little abuse of notation, let $w^*_k$ denote the solution of $$\bar\delta_{w[k]}(x[k],w^*_k)=0$$ in SGS under fixed policy $\pi_{w[k]}$.

According to the inequality $e^{kx}>\sum_{q=0}^{\infty}\frac{k^q x^q}{q!}$ and note that there is $\phi(x,a)=\mathcal{O}(x^2)$, the boundedness of $g_p(x)$ defined in Lemma~\ref{lmm: A bounded} implies the boundedness of $A_{w^*[k]},b_{w^*[k]}$. With the constructed step size $\{\alpha_k\}$, we have $\|A_{w^*[k]}\|<B_{\phi}$, $\|b_{w^*[k]}\|<B_r$, and $\mathsf E[\alpha^2_k\|\delta_{w[k]}(x,w)-\bar\delta_{w[k]}(x,w)\|^2]\leq\alpha^2_kB^2_\delta$, where $B_\phi,B_r, B_\delta$ are finite positive constants.

Following \cite{zhang2023convergence}, considering the weights update equation, the auxiliary sequence $\{u[k]\}$ is defined as 
\begin{align*}
    &u[0]:=w[0],\\
    &u[k+1]:=\Gamma(u[k])+\alpha_k\delta_{w[k]}(x[k],\Gamma(u[k])).
\end{align*}
Since the $\hat Q$ and policy $\pi$ are both linearly related to $w$, we have $w[k]=\Gamma(u[k]),\forall k\in\mathbb{Z}_{\geq 0}$. Let $y'=u[k+1]-w^*_{k+1},~y=w[k]-w^*_k$, then we have
\begin{align*}
    &\mathsf E[\frac{1}{2}\|y'\|^2]=\mathsf E\Big[\frac{1}{2}\|y\|^2\Big]\\
    &\quad+\underbrace{\mathsf E\Big[\Big\langle y,\quad\alpha_k\delta_{w[k]}(x[k],w[k])+w^*_k-w^*_{k+1}\Big\rangle\Big]}_{T_2}\\
    &\quad+\underbrace{\mathsf E\Big[\frac{1}{2}\|\alpha_k\delta_{w[k]}(x[k],w[k])+w^*_k-w^*_{k+1}\|^2\Big]}_{T_3},
\end{align*}
where $\langle\cdot,\cdot\rangle$ is defined as section~\ref{subsec: sgs algorithm}.
The second item can be rewritten as
\begin{align*}
    T_2&=\Big\langle y,\mathsf E\Big[\underbrace{\alpha_k\delta_{w[k]}(x[k],w[k])}_{T_{21}}\Big]\Big\rangle+\langle y,w^*_k-w^*_{k+1}\rangle.
\end{align*}
For $T_3$, we have
\begin{align*}
    T_3 &= \mathsf E\Big[\frac{1}{2}\|w^*_k-w^*_{k+1}\|^2\Big]+\mathsf E\Big[\Big\langle w^*_k-w^*_{k+1}, T_{21}\Big\rangle\Big]\\
    &\quad+\mathsf E\Big[\frac{1}{2}\|T_{21}\|^2\Big].
\end{align*}

Now we are ready to analyze the boundedness of each item. 
Note that $\alpha_k\not=0$ only when $x[k]\in\tilde{\mathcal{X}}$, with \cite[Theorem 4.4., Lemma C.5., Lemma D.10.]{zhang2023convergence} and the analysis of $\delta_{w[k]}$ and $w^*_k$, we have
\begin{align*}
    \alpha_k\|\delta_{w[k]}(x[k],w)\|\leq \alpha_k(L_F\|w\|+U_F),
\end{align*}
and
\begin{align*}
    \|w^*_{w}-w^*_{w'}\|&\leq\underbrace{(B^2_AB_{\phi}L_{DP}+B_AL_D)L_{\pi}B_r}_{L_w}\|w-w'\|\\
&\leq \alpha_kL_w(U_F+L_FC_\Gamma+C_\Gamma)=\alpha_kL_{BW},
\end{align*}
where $B_A, U_F, L_F$ are finite positive constants that $B_A>\sup_{w}\|A^{-1}_{w^*}\|,~\alpha_kU_F\geq\|\alpha_kb_k\|,~L_F\geq((1+\gamma)(B_{\tilde{\mathcal{X}}}+1)^2+1)$, and $L_D,L_{DP}$ are Lipschitz constants of state distribution and transition probability. We have
\begin{align*}
    T_{21}&=\alpha_k\bar\delta_{w[k]}(\cdot)+\alpha_k\delta_{w[k]}(\cdot)-\alpha_k\bar\delta_{w[k]}(\cdot),
\end{align*}
where $(\cdot)$ is short for $(x[k],w[k])$. By leveraging \cite[Lemma D.2.]{zhang2023convergence}, suppose that $k$ is sufficient large, we have $\|\alpha_k\bar\delta_{w[k]}(\cdot)\|\leq(k_\alpha-1)\|y\|,~k_\alpha=\sqrt{1-B_\phi\alpha_k}\in(0,1)$. Then we have
\begin{align*}
    T_2\leq
    \mathsf E[\Big((k_\alpha-1)\|y\|+\alpha_kB_\delta+\alpha_kL_{BW}\Big)\|y\|],
\end{align*} where $k_\alpha=\sqrt{1-B_\phi\alpha_k}\in(0,1)$. Analogously, we have
\begin{align*}
    T_3&\leq\mathsf E[\frac{1}{2}(k_\alpha-1)^2\|y\|^2+\alpha_k(k_\alpha-1)(B_\delta+L_{BW})\|y\|\\
    &+\alpha_k^2(\frac{1}{2}B^2_\delta+B_\delta L_{BW}+\frac{1}{2}L^2_{BW})].
\end{align*}

According to the above analysis, we have 
\begin{align*}
    &\mathsf E[\frac{1}{2}\|u[k+1]-w^*_{k+1}\|^2]\\
    &=\frac{1}{2}\mathsf E\Big[\Big(k_\alpha\|w[k]-w^*_k\|+\alpha_k(L_{BW}+B_\delta)\Big)^2\Big].
\end{align*}
Then we have
$$
z_{k+1}\leq k_\alpha z_k+\alpha_k(L_{BW}+B_\delta),
$$where $z_{k+1}=\sqrt{\mathsf E[\|u[k+1]-w^*_{k+1}\|^2]}$.
Since $k_\alpha\in(0,1)$ and $\|w[k]-w^*_k\|\leq\|u[k]-w^*_k\|,$ by iteration, we have
\begin{align*}
    z_{k+1}&=\sqrt{\prod^k\nolimits_{\ell=k_0}(1-\alpha_\ell B_\phi)}\cdot z_{k_0}\\
    &+(L_{BW}+B_\delta)\sum^k\nolimits_{\ell=k_0}\alpha_\ell\sqrt{\prod^k\nolimits_{j=\ell}(1-\alpha_jB_\phi)}.
\end{align*}
Note that $\{\alpha_k\}$ is constrained by \eqref{eq_sumk=0} and the inequality $1-x\leq e^{-x}$ holds, we have
\begin{align*}
    z_{k+1}\leq\frac{B_\delta +L_{BW}}{B_\phi}.
\end{align*}
Considering the relationship between the policy and weight vector, we have
\begin{align*}
    \mathsf E[\|w[k]-w^*\|]&\leq\mathsf E[\|w[k]-w^*_k\|]+\mathsf E[\|w^*_k-w^*_{w^*}\|]\\
    &\leq \mathsf E[\|w[k]-w^*_k\|]+L_w\mathsf E[\|w[k]-w^*\|].
\end{align*}
When the immediate cost $c$ is constrained such that $L_w<1$, we have
\begin{align*}
    \mathsf E[\|w[k]-w^*\|]&\leq\frac{1}{1-L_w}\mathsf E[\|w[k]-w^*_k\|]\leq\frac{1}{1-L_w} z_t.
\end{align*}
Then finally we can conclude that
\begin{align*}
    \mathsf E[\|w[k]-w^*\|]\leq\frac{B_\delta +L_{BW}}{(1-L_w)B_\phi},
\end{align*} 
which yields \eqref{equa: algo convergent}.
\section{experiments}\label{sec_exp}
To evaluate the performance of the semi-gradient SARSA (SGS) algorithm with weighted shortest queue (WSQ) policy, we consider two benchmarks:
\begin{itemize}
    \item Neural network (NN)-WSQ: We constructed a NN for approximation of $Q(x,a)$. The algorithm is similar to Algorithm~\ref{algo: SGS-WSQP}, except that the weights update is replaced by NN update with adaptive moment estimation (Adam) algorithm \cite{Kingma2014AdamAM}. Specifically, the NN has two fully connected layer with a rectified linear unit (ReLU) activation function. The loss function is the mean square error between the one-step predicted and calculated state-action-value. Since an exact optimal policy of the original MDP is not readily available, the policy computed by NN is used as an approximate optimal policy.
    \item Join the shortest queue (JSQ) policy: For routing decisions under JSQ policy, we simply select the path with the shortest queue length, that is $a_{JSQ}=\arg \min_{n\leq N} x_n$.
\end{itemize}
Consider the network in Fig.~\ref{fig:parallelQN} with three parallel servers. Suppose that the service rate $\mu_1 = 0.5,~\mu_2 = 2.5,~\mu_3=5$ and the arrival rate $\lambda = 2$, all in unit sec$^{-1}$. The WSQ policy temperature parameter is set as $\iota=0.01$. For SGS algorithm, we initialize the weight as $w_1=w_2=w_3= 0.5$. For simulation, a discrete time step of 0.1 sec is used. All experiments were implemented in Google Colab \cite{colab}, using Intel(R) Xeon(R) CPU with 12.7GB memory. We trained SGS for $10^6$ episodes, NN for $4\times10^6$ episodes, and evaluate them for $10^6$ episodes each, and the result are as follows.

For the performance, the weight of SGS converges to $w = [0.60\ 0.49\ 0.15]$, which means the weight of a slower server is higher. Our WSQ policy is not restricted to non-idling conditions, since $w_1(2\times0+1)>w_3(2\times1+1)$, the server with higher service rate (i.e., server 3) is more likely to be selected, even the slower server (i.e., server 1) is empty. 

\begin{figure}[htbp]
        \centerline{\includegraphics[width=0.7\linewidth]{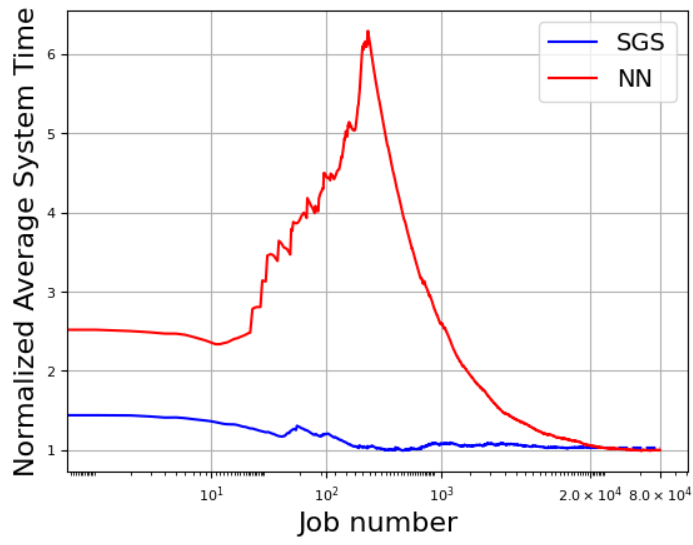}}
    \caption{The performance compare between SGS and NN.}
    \label{fig:comp_performance}
\end{figure}

\begin{table}[htbp]
    \centering
    \caption{Average system times of various schemes.}
    \label{table:ave_job_T}
    \begin{tabular}{cccc}
        \toprule  
        Algorithm&Normalized Average System Time\\
        \midrule  
        Neural network (NN) &1.00\\
        Semi-gradient SARSA (SGS)&1.08\\
        Join the shortest queue (JSQ)&2.78\\
        \bottomrule  
    \end{tabular}
\end{table}

Table \ref{table:ave_job_T} lists the normalized average system time results under various methods. The results are generated from test simulations of $10^5$ sec. Although NN performs better in long terms of learning as expected, SGS performs better with just a few number of iterations as demonstrated in Fig. \ref{fig:comp_performance}.

The job will spend more time going through the queuing network under JSQ policy at the average of $ 0.8581$ sec. For WSQ, though the implementation efficiency of SGS is slightly worse than NN, SGS gives the best trade-off between computational efficiency and implementation efficiency: the average system time of NN and SGS are $0.3078$ sec and $0.3318$ sec. More importantly, SGS algorithm theoretically ensures the convergence of the optimal routing decision, while NN might be diverge and let alone the existence of the optimal decision.
\section{conclusion}\label{sec_con}
In this paper, we propose a semi-gradient SARSA(0) (SGS) algorithm with linear value function approximation for dynamic routing over parallel servers. We extend the analysis of SGS to infinite state space and show that the convergence of the weight vector in SGS is coupled with the convergence of the traffic state, and the joint convergence is guaranteed if and only if the paralle service system is stabilizable. Specifically, the approximator is used as Lyapunov function for traffic state stability analysis; and the constraint and convergence analysis of weight vector is based on stochastic approximation theory. Besides, our analysis can be extended to polynomial approximator with higher order. We compare the proposed SGS algorithm with a neural network-based algorithm and show that our algorithm converges faster with a higher computationally efficiency and an insignificant optimality gap. Possible future work includes (i) extension of the joint convergence result as well as SGS algorithm to a general service network and (ii) the analysis of fixed-point convergence condition.
\bibliographystyle{IEEEtran}
\bibliography{main}
\end{document}